\documentclass{article}

\makeatletter
\newcommand{\chapterauthor}[1]{%
  {\parindent0pt\vspace*{0pt}%
  \linespread{1.0}\normalsize\scshape#1%
  \par\nobreak\vspace*{15pt}}
  \@afterheading%
}
\makeatother

\usepackage[margin=1in]{geometry}
\usepackage{parcolumns,lipsum}
\usepackage{verbatim}
\usepackage{subcaption}
\usepackage{url}
\usepackage{xcolor}

\title{Enrico Fermi: a little-known conference dating back to the early years of solid-state physics}
\author{Emanuele Goldoni, Ledo Stefanini\\ \textit{Accademia Nazionale Virgiliana}\\ \textit{Mantova, Italy}}
\begin{document}
\maketitle

\begin{abstract}
In April 1925, Enrico Fermi was only 23 years old and he had graduated less than three years earlier. Despite his age, Fermi was invited at the University of Rome by Federigo Enriques to give several lectures on the recent advances in physics at the Italian Mathematics Seminar. 
Enriques was also the editor of ``Periodico di Matematiche'', an important Italian journal concerned primarily with secondary school math teachers. Some of the articles published by ``Periodico di Matematiche'' correspond to the lectures which Fermi gave during seminars and meetings of Italian mathematicians. Here we propose an English translation for one of these articles, entitled ``Sopra la teoria cinetica dei corpi solidi'' and discussing the kinetic theory of solids. We believe that this Fermi's work offers a twofold interesting perspective: on one side, it allows us to explore the problems related to the structure of matter of emerging at the time as seen by the eyes of a bright, young physicist. On the other hand, it shows us the undoubtedly high scientific level of the articles read in Italy by the secondary school math teachers in the Twenties.
\end{abstract}

\section*{Introduction}
In the Fall of 1924, thanks to the interest of the mathematician Vito Volterra, Enrico Fermi received a scholarship and spent a few months in Leiden at the institute directed by Paul Ehrenfest. In early January 1925, supported by the distinguished scientist and powerful senator Orso Mario Corbino, the young physicist received a position at the University of Florence, where he taught Mathematical Physics and Analytical Mechanics to engineers. At that time, Fermi was only 23 years old and he had graduated less than three years earlier.

In April 1925, Fermi was invited to give several lectures on the recent advances in physics at the Italian Mathematics Seminar at the University of Rome. 
The activities of the Mathematics Seminar were directed by Federigo Enriques, who had held the chair of Higher Mathematics and Higher Geometry at the University of Rome since 1922. Enriques has been one of the greatest figures of the Italian school of Mathematics and his scientific interests were quite broad -- it was Enriques himself who promoted a series of lectures given at the University of Bologna by Albert Einstein in October 1921.

Enriques was also the editor of ``Periodico di Matematiche'', an important Italian journal targeted primarily to secondary school teachers of mathematics. Enriques was not a physicist but he strongly believed that the journal should deal not only with mathematics but also, at an elementary level, with the most modern subjects of physics. 
Some articles published by ``Periodico di Matematiche'' correspond to the lectures which Fermi gave in the 1920s during the seminars and the meetings of Italian mathematicians \cite{Fermi:1925:STC, Fermi:1925:SPD, Fermi:1926:PCN, Fermi:1928:PMC}. It can be presumed, therefore, that these articles were solicited from Fermi by Enriques, who held a very high opinion of the young physicist.

One of the transcriptions of Fermi's lessons is entitled ``Sopra la teoria cinetica dei corpi solidi'' ("On the kinetic theory of solids"): it appeared originally in \cite{Fermi:1925:STC} and was later republished in \cite{Fermi:1962:collected1} with the different title ``Sopra la teoria dei corpi solidi'' (without the word \textit{kinetic}). This article offers a twofold interesting perspective: on one side, it allow us to explore the problems related to the structure of matter emerging at the time as seen by the eyes of a bright young physicist (as well as future Nobel laureate). On the other hand, this work shows us the undoubtedly high scientific level of the publications read in Italy by the secondary school math teachers in the Twenties, mainly thanks to efforts of Federigo Enriques and his colleagues.

In this article we propose an English translation for ``Sopra la teoria cinetica dei corpi solidi'' in order to make the full text available to a broader public. About a century has passed since its publication: we have tried to keep the style and the vocabulary as faithful to the original text as possible, although the results could seem a bit far from modern scientific publications. Nonetheless, we hope that our effort will allow the readers to see the world of matter through the eyes of Enrico Fermi, as some lucky Italian mathematicians had the opportunity on that distant April 18, 1925.

\setcounter{equation}{0}
\section*{On the cinectic theory of solids}
\chapterauthor{Enrico Fermi (Author), Emanuele Goldoni and Ledo Stefanini (Translators)}

From a kinetic point of view, what distinguishes a solid from a liquid consists essentially in the fact that, while the atoms of the latter in their thermal motion can move considerably far from the position they initially occupied, those of the former oscillate continuously around an equilibrium position. 
These different behaviors are obviously due to the different relationship that exists in the two cases between the kinetic energy of the thermal motion of the atoms, and the energy of the forces that tend to keep them in their equilibrium positions. 

The fundamental problems of the kinetic theory of solids can be fundamentally reduced to two questions. The first one aims at the determination of the thermal motion of atoms, admitting (tentatively) that the forces acting between them are quasi-elastic; on the other hand, the object of the second problem is the determination of the actual nature of these forces.

Initially, we will deal with the former problem, and specifically we will determine the energy of thermal motions – and thus also the specific heat capacity – as a function of temperature. This problem, which historically was addressed first as well, is especially well suited to show the successive stages in the development of the theory, and how this theory had to be refined along the way to account for new experimental facts as they were discovered.

The specific heat capacity of solids varies with the temperature. There is, however, a temperature range -- which includes, for most solid elements, the ordinary temperature -- in which the specific heat capacity remains fairly approximately constant. The first measurements were naturally made in this range and led to the discovery of the law of Dulong and Petit, which states, as is well known, that the specific heat capacity of solid elements is inversely proportional to their atomic weight -- in other words, that the specific heat capacity is equal for all elements if we refer to the gram atom instead of referring to the unit of mass.

This law has very few exceptions when making measurements at ordinary temperatures and it found an easy theoretical interpretation. In fact, let's consider $N$ atoms of a body which, to a first approximation, we will regard as material points; they will have a total of $3N$ degrees of freedom. In accordance with the principles of statistical mechanics, to each degree of freedom corresponds the average kinetic energy $kT/2$, where $T$ is the absolute temperature and $k$ the Boltzmann constant; therefore, the kinetic energy of our $N$ atoms will be $3NkT/2$. Since the forces acting between them are elastic forces to a first approximation, we have that on average the kinetic energy is equal to the potential energy.

Hence, the total thermal energy of our body will be
\begin{equation} W = 3NkT \end{equation}

From this expression of the energy, we can immediately obtain the expression of specific heat capacity, which is given by
\[ c = \frac{dW}{dt} = 3Nk \]

According to this theory, is thus equal for all elements and independent of the temperature. Its value well-matches the constant of Dulong and Petit's law in numerical terms too.
However, later experiments showed that Dulong and Petit's law completely loses its validity for very low temperatures. At these temperatures, the specific heat capacity ceases to be constant and it decreases towards zero as the temperature approaches absolute zero.

Einstein was the first who gave a theoretical interpretation of this result. He observed that the atoms of a body are equivalent to oscillators since they are bound together by quasi-elastic forces -- to a first approximation, we will consider them be equivalent to $3N$ oscillators, all having the same frequency $\nu$. According to quantum law, the equipartition theorem loses its validity for high-frequency oscillators; the expression of the average energy of an oscillator is not $kT$, as in classical statistical mechanics, but has instead the value:
\[ \frac{h \nu}{e^{\frac{h \nu}{kT}}-1} \]

Therefore, the average energy of the $3N$ oscillators - equivalent to our body - will be
\begin{equation} W = \frac{3Nh\nu}{e^{\frac{h \nu}{kT}}-1} \end{equation}

This formula and the formula for the specific heat capacity that can be deduced from it, both summarily agree with experience, especially for temperatures that are not extremely low. Similarly, the values of $\nu$ calculated from the trend of the specific heat capacities, correspond quite well with the frequencies of the so-called Reststrahlen, which are light radiations located in the extreme infrared arising precisely from the vibrations of the atoms of the body.

However, Einstein's theory too proved insufficient for temperatures very close to absolute zero; in fact, for extremely low temperatures the specific heat capacity of solids tends to zero as $T^3$, while it should cancel out much more rapidly according to Einstein's theory.

To correct this last divergence, it was necessary to modify an overly simplistic assumption in Einstein's theory. This theory, in fact, assumes that all characteristic frequencies of the solid are equal. To the contrary, the elastic system consisting of all the atoms of the body will be capable of vibrating at a great number of frequencies; therefore, all these characteristic frequencies must be taken into account separately when calculating the thermal energy of the solid.

A first method for calculating such frequencies was devised by Debye: although it made use of an unsatisfactory hypothesis at first glance, it nevertheless provided very good results. Later, Debye's method was perfected and completed by Born and Karman, who also devised a way of justifying the assumptions made by Debye. The method used by Debye is based on the following principle: first of all, the solid is considered as a homogeneous and continuous elastic medium, ignoring to a first approximation its discontinuous structure. Then, if we consider a piece of this solid, it will be capable of vibrating at certain characteristic frequencies depending on its shape, size and elastic properties, in the same way that a taut string can vibrate at frequencies that depend on its length and tension. Of course, for the problem at issue, it is not necessary to determine all these frequencies in detail: it is sufficient to know their distribution, that is, to have an asymptotic expression for the number of characteristic frequencies that fall in an assigned frequency range.

Now we can find that the number of eigenfrequencies between $\nu$ and $\nu + d\nu$ is given by $\frac{4\pi V}{v^3}\nu^2d\nu$ in a system of vibrations propagating with velocity $v$ in a body of volume $V$. Moreover, since there are three propagating systems in the case of an elastic solid -- two transverse with velocity $v_1$ and one longitudinal with velocity $v_2$ -- we can conclude that the number of eigenfrequencies between $\nu$ and $\nu + d\nu$ is given by $4\pi V \left({\frac{2}{v_1^3} + \frac{1}{v_2^3}}\right)\nu^2 d\nu$. Attributing to each of these frequencies the average energy $ \frac{h \nu}{e^{\frac{h \nu}{kT}}-1}$, we can express the thermal energy of our whole solid as
\begin{equation} W = 4\pi V \left({\frac{2}{v_1^3} + \frac{1}{v_2^3}}\right) \int{\frac{h \nu^3 d\nu}{e^{\frac{h \nu}{kT}}-1}} .\end{equation}

If the assumptions we made considering our body as a continuous elastic body were accurate, clearly the former integral would have to be extended between zero and infinity because it would possible to propagate elastic vibrations at arbitrarily large frequencies.

However, it is obvious that this cannot happen in our case: while the continuous body is capable of vibrating with infinite characteristic frequencies, the real discontinuous body, on the other hand, has only $3N$ degrees of freedom. Facing this difficulty, Debye used an artifice which was very unsatisfactory in principle but led to brilliant results; moreover, it was justified later -- at least partly -- by the more refined theory of Born and Karman. Debye admitted that in a real elastic body, vibrations of frequencies higher than a certain limit $\nu_{max}$ could not propagate. He calculated this limit so that the total number of characteristic frequencies lower than this limit, expressed as
\[ 4\pi V \left({\frac{2}{v_1^3} + \frac{1}{v_2^3}}\right) \int\limits_0^{\nu_{max}}{\nu^2 d\nu} \]
would be equal to $3N$, which is the number of degrees of freedom of the body.

Under this assumption, the upper limit of the integral (3) becomes $\nu_{max}$ and we obtain an expression for thermal energy -- and thus also for the specific heat capacity -- which is in excellent agreement with experience. We also observe that the alignment between this theory and the experience is really striking because the Debye's formula contains the elastic constants of the body as its only parameters, which can be determined by direct elastic measurements completely independent of specific heat capacity measurements. 

As we have already mentioned, a further refinement to the theory of specific heat capacities was brought by Born and Karman. Starting from the same concept as Debye -- that is to say to calculate the thermal energy of the solid as the sum of the energies of many oscillators of frequencies equal to the characteristic frequencies of the body -- they applied a much more correct method for the calculation of these frequencies considering the body as it actually is, namely a solid consisting of a lattice of atoms. Without going into the details of the calculus of an elastic system consisting of a lattice of masses exerting forces between them proportional to their relative displacements, we want to use a simplified example to show the influence that the discontinuous structure of the body has over the propagation of elastic waves.

Hence, we will study the propagation of an elastic perturbation in a row of points of equal mass, spaced apart by a constant interval $a$. We will assume that over each atom only the two closest atoms will act, and that the force exerted between two adjacent atoms is proportional to their relative displacement.  Then, we suppose that a sinusoidal elastic wave, of frequency $\nu$ and velocity $v$, propagates along our row. The displacement of a particle can be written as
\[ \delta = A \sin {2\pi\nu\left( t - \frac{x}{v} \right)} \]
where $x$ represents the abscissa of the particle, which is given by $n a$, being $n$ an integer. We aim to find the relationship between $\nu$ and $v$; we will show precisely that -- unlike what would happen in an elastic bar with a continuous structure -- the speed of propagation is not constant but depends on the frequency.

The displacement of the particle for $n=1$ is given by $A \sin {2\pi\nu\left( t - \frac{a}{v} \right)}$; instead, that of the particle $n=0$ is given by $A \sin {2\pi\nu t}$. The force acting between these two particles, which by assumption is proportional to their relative displacement, can therefore be written as
\[ k A \left\{ \sin{2\pi\nu\left( t - \frac{a}{v} \right)} - \sin{2\pi \nu t} \right\} \]
being $k$ the coefficient of proportionality. Due to the action of the particle for which $n=-1$, on the particle $n=0$ another force will also act. This force will be similarly expressed by $k A \left\{ \sin{2\pi\nu\left( t + \frac{a}{v} \right)} - \sin{2\pi \nu t} \right\}$

Writing that the product of the mass $m$ of the particle for which $n=0$, by its acceleration $-4 \pi^2\nu^2 A \sin{2\pi \nu t}$ is equal to the sum of the forces acting on it, we find
\[ -4 \pi^2\nu^2 m A \sin{2\pi \nu t} = k A \left\{ \sin{2\pi\nu\left( t - \frac{a}{v} \right)} + \sin{2\pi\nu\left( t + \frac{a}{v} \right)} - 2\sin{2\pi \nu t} \right\} = \]
\[ = -2 k A \sin 2\pi \nu t \left\{ 1 - \cos{\frac{2\pi \nu a}{v}} \right\} = -4 k A \sin{2\pi \nu t \sin^2{\frac{\pi \nu a}{v}}}\]
From this equation we derive, with simple reductions
\begin{equation}
\nu = \frac{1}{\pi} \sqrt{k \over m} \sin{\frac{\pi \nu a}{v}}.
\end{equation}

We have found precisely the relationship between propagation speed and frequency we were looking for. The case of the propagation in a continuous medium can be obtained considering the borderline case $a=0$. Here, substituting the sine we see that $\nu$ disappears from the previous formula; thus the propagation velocity $v$ turns out to be independent of the frequency, as indeed it must be in a continuous-structure propagating media.

We also observe that, since the sine can never be greater than 1, from (4) we get
\[ \nu \leq \frac{1}{\pi} \sqrt{k \over m} \]
which tells us that no propagation of elastic waves of frequency higher than $\frac{1}{\pi} \sqrt{k \over m}$ can occur in our row of particles. In a sense, this fact justifies the introduction of the limiting frequency arbitrarily made by Debye.

Let's now move on to briefly mention how we can interpret some of the main phenomena of the physics of solids using from the point of view of the theory we have just seen. 

If the forces acting between the atoms of the body were strictly elastic, the average position occupied by each atom should not shift due to thermal motion. However, in reality the interatomic forces are only approximately elastic. The consequence of this fact is that the moving atom also moves on average from the position it would occupy if it were stationary, so that due to thermal motion the distance between the atoms varies. This is what leads to thermal expansion. To see schematically how this happens, we consider a single point, attracted by a quasi-elastic force to a position of equilibrium. To simplify, we will assume that our point is moving along a line, where $x = 0$ is the equilibrium position on the abscissa. We consider $U = kx^2 + hx^3$ as the expression of the potential energy of the quasi-elastic force; if the cubic term were missing the force would be exactly elastic, therefore we will assume that the cubic term is very small in comparison with the quadratic term. The force acting over our particle will be expressed by $-\frac{\partial u}{\partial x} = -2kx - 3 hx^2$; the equation of motion will be
\[ mx'' = -2kx -3hx^2 .\]

Let's take the mean values of the two terms in the previous equation. If we assume that the velocity remains finite, we will obviously have $\overline{x''} = 0$. Therefore
\[ -2k\overline{x} -3h\overline{x^2} = 0 \]
i.e. 
\begin{equation} \overline{x} = - \frac{3}{2} \frac{h}{k} \overline{x^2} .\end{equation}
It can be seen, therefore, that the non-perfect elasticity of the bond makes the mean value of the abscissa different from zero, that is, it produces a shift in the mean position of the particle. It is clear how this kind of reasoning could be extended to the case of an atomic lattice, and it is understandable why  thermal agitation has the effect of expanding substances. We observe also that from (5) that $\overline{x}$ results proportional to $\overline{x^2}$, that is, in first approximation to the energy of motion. Therefore, we have that the thermal expansion will be proportional to the thermal energy of the body. And this actually occurs, because experience shows that both the specific heat capacity and the coefficient of thermal expansion have an entirely similar trend as a function of temperature.

The non-complete elasticity of the atom-to-atom bond in a solid plays an important role for explaining the thermal conduction of insulators, pyroelectricity, and several other phenomena. However, we are running short on time and so we cannot deal with these topics.

Instead, I want to quickly mention the studies that have been made on the nature of the forces acting between two atoms. This study, mainly due to the work of Born and his school, has so far led to definitive results only for the case of substances with a salt-like structure. This depends on the fact that in non-saline bodies we have, probability, electrons migrating from one atom to another describing entangled orbits; it follows in this case an extraordinary complication of the interatomic forces, which until now have remained indecipherable.

Crystal lattices of salts, on the other hand, consist of ions which, even as constituents of the crystal, always retain their individuality. For example, let's  consider a piece of halite: its crystal lattice consists of positive ions of sodium arranged alternately with negative ions of chlorine. A sodium ion consists of a positive nucleus, having 11 times the charge of the electron, surrounded by 10 negative electrons: the compound has a negative charge of a magnitude equal to that of the electron. At a great distance from the ion, the electric force it produces will be -- to a great approximation -- equal to that of a positive point charge; on the other hand, its electric potential will also depend strongly on its complex structure when very close to the ion. In contrast, the negative chlorine ion consists of a positive nucleus, having a charge 17 times that of the electron, surrounded by 18 electrons. Therefore, at a great distance, its electric potential will be that of a negative point charge; on the other hand, similarly to the sodium ion, the influence of its complex structure will be significant when we move very close to it. According to Coulomb's law, if we now suppose that a chlorine ion and a sodium ion are placed in the presence of each other, we would see that they will attract each other as two point charges, one positive and one negative, as long as they are very far apart. When the two ions come very close, the force acting between them will depend fundamentally on the arrangement of the electrons. It is not possible to make an exact calculation of the force acting between the two ions in this case, owing to our ignorance of the details of their electronic structure; nevertheless, we can predict that when the electrons surrounding the two ions come very close together the repulsive force between them will eventually prevail at some point so that in the immediate vicinity the two ions will repel each other. The force acting between the two ions can thus be represented as the sum of an attractive force, inversely proportional to the square of the distance, and a repulsive force that acquires very large values when the two ions are very close to each other and decreases very rapidly with distance.

A brilliant experimental confirmation of these views occurs in the calculation of the energy holding together a lattice of ions. The experiments of Laue, Bragg and others over the diffraction phenomena of Röntgen rays have indeed made it possible in recent times to determine exactly the position of atoms in the crystal lattices of numerous substances. Then, very laborious calculations make it possible to calculate the potential energy of interatomic forces, which is nothing more than the energy that is required to separate the crystal into its constituent ions, and that can also be measured directly. The agreement between the measured and the calculated values of this energy proved to be practically perfect. This confirmation, as well as the success achieved in calculating the compressibility and other elastic constants of salts, have now confirmed beyond any doubt the electrical nature of the forces that hold together the atoms of these substances.

\section*{Acknowledgements}
We would like to thank Davide Cavalca for giving the paper a critical reading and for providing several helpful comments. 



\end{document}